\documentclass[10pt,twoside]{hsqcd}
\usepackage{epsf,amsmath}
\usepackage{graphicx}

\def\ktf {$k_t$-factorization }

\def\a {\epsilon}
\def\g {\gamma}

\def\F  {{\cal F}}
\def\p  {{\cal P}}
\def\J {$J/\psi$ }

\def\U {$\Upsilon$ }
\def\u { \Upsilon  }

\def\qq {$Q\bar{Q}$ }
\def\cpc#1#2#3  {{Computer\ Phys.\ Comm.\ }  {\bf#1}, #2 (#3)}
\def\err#1#2#3  {{\it Erratum }              {\bf#1}, #2 (#3)}
\def\epjc#1#2#3 {{Eur. Phys. J. C }          {\bf#1}, #2 (#3)}
\def\dum#1#2#3  {{~}                         {\bf#1}, #2 (#3)}
\def\ib#1#2#3   {{\it ibid. }                {\bf#1}, #2 (#3)}
\def\jcp#1#2#3  {{J.\ Comp.\ Phys.\ }        {\bf#1}, #2 (#3)}
\def\jhep#1#2#3 {{JHEP }                     {\bf#1}, #2 (#3)}
\def\ijmp#1#2#3 {{Int.\ J.\ Mod.\ Phys.\ }   {\bf#1}, #2 (#3)}
\def\jpg#1#2#3  {{J.\ Phys.\ G }             {\bf#1}, #2 (#3)}
\def\mpl#1#2#3  {{Mod.\ Phys.\ Lett.\ }      {\bf#1}, #2 (#3)}
\def\ncim#1#2#3 {{Nuovo Cimento }            {\bf#1}, #2 (#3)}
\def\np#1#2#3   {{Nucl.\ Phys.\ }            {\bf#1}, #2 (#3)}
\def\npb#1#2#3  {{Nucl.\ Phys.\ B}           {\bf#1}, #2 (#3)}
\def\pan#1#2#3  {{Phys.\ At.\ Nuclei }       {\bf#1}, #2 (#3)}
\def\plb#1#2#3  {{Phys.\ Lett.\ B }          {\bf#1}, #2 (#3)}
\def\prep#1#2#3 {{Phys.\ Rep.\ }             {\bf#1}, #2 (#3)}
\def\prd#1#2#3  {{Phys.\ Rev.\ D }           {\bf#1}, #2 (#3)}
\def\prl#1#2#3  {{Phys.\ Rev.\ Lett.\ }      {\bf#1}, #2 (#3)}
\def\ptp#1#2#3  {{Prog.\ Theor.\ Phys.\ }    {\bf#1}, #2 (#3)}
\def\ppnp#1#2#3 {{Prog.\ Part.\ Nucl.\ Phys.\ } {\bf#1}, #2 (#3)}
\def\ps#1#2#3   {{Physica Scripta }          {\bf#1}, #2 (#3)}
\def\rmp#1#2#3  {{Rev.\ Mod.\ Phys.\ }       {\bf#1}, #2 (#3)}
\def\rpp#1#2#3  {{Rep.\ Prog.\ Phys.\ }      {\bf#1}, #2 (#3)}
\def\sa#1#2#3   {{Sci. Acta}                 {\bf#1}, #2 (#3)}
\def\sjnp#1#2#3 {{Sov.\ J.\ Nucl.\ Phys.\ }  {\bf#1}, #2 (#3)}
\def\spj#1#2#3  {{Sov.\ Phys.\ JETP }        {\bf#1}, #2 (#3)}
\def\spjl#1#2#3 {{Sov.\ JETP Lett.\ }        {\bf#1}, #2 (#3)}
\def\spu#1#2#3  {{Sov.\ Phys.-Usp.\ }        {\bf#1}, #2 (#3)}
\def\yaf#1#2#3  {{Yad.\ Fiz.\ }              {\bf#1}, #2 (#3)}
\def\zp#1#2#3   {{Zeit.\ Phys.\ }            {\bf#1}, #2 (#3)}
\def\zpc#1#2#3  {{Z.\ Phys.\ C }             {\bf#1}, #2 (#3)}
\def\etal {{\it et al. }}

\begin{document}

\title{UPSILONIUM POLARIZATION AS A TOUCHSTONE IN \\
UNDERSTANDING THE PARTON DYNAMICS IN QCD
\thanks{This work is
supported by the FASI of RF (Grant No.NS-1856.2008.2), the RFBR foundation (Grant No.
08-02-00896-a) and DESY Directorate in the framework of Moscow--DESY project on MC implementation for HERA--LHC }}

\author{\underline{N.~P.~ZOTOV} and S.~P. BARANOV$^1$\\ \\
SINP, Moscow State University, 
Moscow 119991, Russia\\
E-mail: zotov@theory.sinp.msu.ru\\
1 - Lebedev Physics Institute,
Moscow 119991. Russia}

\maketitle

\begin{abstract}
\noindent 
In the framework of the \ktf approach, the production of \U mesons 
at the Fermilab Tevatron is considered, 
and the comparisions of 
calculated $p_T$-distributions and
spin alignment parameter $\alpha$  with the D0 
experimental data are shown.
We argue that measuring the double cross section and  
the polarization of upsilonium
states can serve as a crucial test discriminating two competing
theoretical approaches to parton dynamics in QCD.
\end{abstract}



\markboth{\large \sl \underline{N.P. Zotov} \& S. P. Baranov
\hspace*{2cm} HSQCD 2008} {\large \sl \hspace*{1cm} TEMPLATE FOR THE
HSQCD 2008 PROCEEDINGS}

\section{Introduction} 
Nowadays, the production of heavy quarkonium states at high energies 
is under intense theoretical and experimental study \cite{ref1,ref2}.
The production mechanism involves the physics of both short and long
distances, and so, appeals to both perturbative and nonperturbative 
methods of QCD. The creation of a heavy quark pair \qq proceeds via 
the photon-gluon or gluon-gluon fusion (respectively, in $ep$ and $pp$ 
collisions) referring to small distances of the order of $1/(2m_Q)$, 
while the formation of the colorless final state refers to longer 
distances of the order of $1/[m_Q\,\alpha_s(m_Q)]$. These distances 
are longer than the distances typical for hard interaction but are yet 
shorter than the ones responsible for hadronization (or confinement).
Consequently, the production of heavy quarkonium states is under 
control of perturbative QCD but, on the other hand, is succeeded by 
nonperturbative emission of soft gluons. This feature gives rise to 
two competing theoretical approaches known in the literature as the 
color-singlet \cite{BaiBer} and color-octet \cite{ChoLei} models.
According to the color-singlet approach, the formation of a colorless
final state takes place already at the level of the hard partonic 
subprocess (which includes the emission of hard gluons when necessary).
In the color-octet model, also known as nonrelativistic QCD (NRQCD),
the formation of a meson starts from a color-octet \qq pair and proceeds 
via the emission of soft nonperturbative gluons.
The former model has a well defined applicability range and has already 
demonstrated its predictive power in describing the \J production at 
HERA, both in the collinear \cite{Kraem} and the \ktf \cite{j_dis} 
approaches. As it was shown in the analysis of recent ZEUS \cite{ZEUS}
data, there is no need in the color-octet contribution, neither in the 
collinear nor in the \ktf approach.

 The numerical estimates of CO 
contributions extracted from the analysis of Tevatron data are at odds 
with the HERA data, especially as far as the inelasticity parameter 
$z=E_{\psi}/E_{\g}$ is concerned \cite{KniZwi}.
In the \ktf approach, the values of the color-octet contributions 
obtained as fits of the Tevatron data appear to be substantially smaller 
than the ones in the collinear scheme, or even can be neglected at all
\cite{j_tev,Teryaev,Chao1,Vasin}.

In the present note we want to compare the predictions of $k_T-$
factorization approach~\cite{bz} with the D0 
experimental data on the $p_T-$ distributions~\cite{D0cs} and the 
polarizaton of $\Upsilon$ (1S) meson states produced at the Tevatron 
energies~\cite{Kuz}.

\section{Numerical results}
In the \ktf approach, the cross section of a physical process is
calculated as a convolution of the partonic cross section $\hat{\sigma}$
and the unintegrated parton distribustion ${\F}_g(x,k_{T}^2,\mu^2)$,
which depend on both the longitudinal momentum fraction $x$ 
and transverse momentum $k_{T}$:
\begin{equation}
  \sigma_{pp} =
  \int {\F}_g(x_1,k_{1T}^2,\mu^2)\,{\F}_g(x_2,k_{2T}^2,\mu^2)\,
  \hat{\sigma}_{gg}(x_1, x_2, k_{1T}^2, k_{2T}^2,...)
  \,dx_1\,dx_2\,dk_{1T}^2\,dk_{2T}^2.
\end{equation}
In accord with the \ktf prescriptions 
\cite{GLR83,Catani,Collins,BFKL},
the off-shell gluon spin density matrix is taken in the form
\begin{equation} \label{epsglu}
 \overline{\a_g^{\mu}\a_g^{*\nu}} =
  p_p^{\mu}p_p^{\nu}x_g^2/|k_{T}|^2 = k_{T}^\mu k_{T}^\nu/|k_{T}|^2.
\end{equation}
In all other respects, our calculations follow the standard Feynman
rules.

In order to estimate the degree of theoretical uncertainty connected
with the choice of unintegrated gluon density, we use two different
parametrizations, which are known to show the largest difference with
each other, namely, the ones proposed in Refs. \cite{GLR83,BFKL} 
and \cite{Bluem}.
In the first case \cite{GLR83}, the unintegrated gluon density is derived
from the ordinary (collinear) density $G(x,\mu^2)$ by differentiating it
with respect to $\mu^2$ and setting $\mu^2=k_T^2$.
Here we use the LO GRV set \cite{GRV98} as the input colinear density.
In the following, this will be referred to as dGRV parametrisation.
The other unintegrated gluon density \cite{Bluem} is obtained as a 
solution of leading 
order BFKL equation \cite{BFKL} in the double-logarithm approximation. 
In the following, this will be referred to as JB parametrisation.

\begin{figure}[!thb]
\hspace*{0mm} 
\begin{minipage}[h]{.49\textwidth}
\includegraphics[height=6.0cm,width=7.0cm,angle=0]{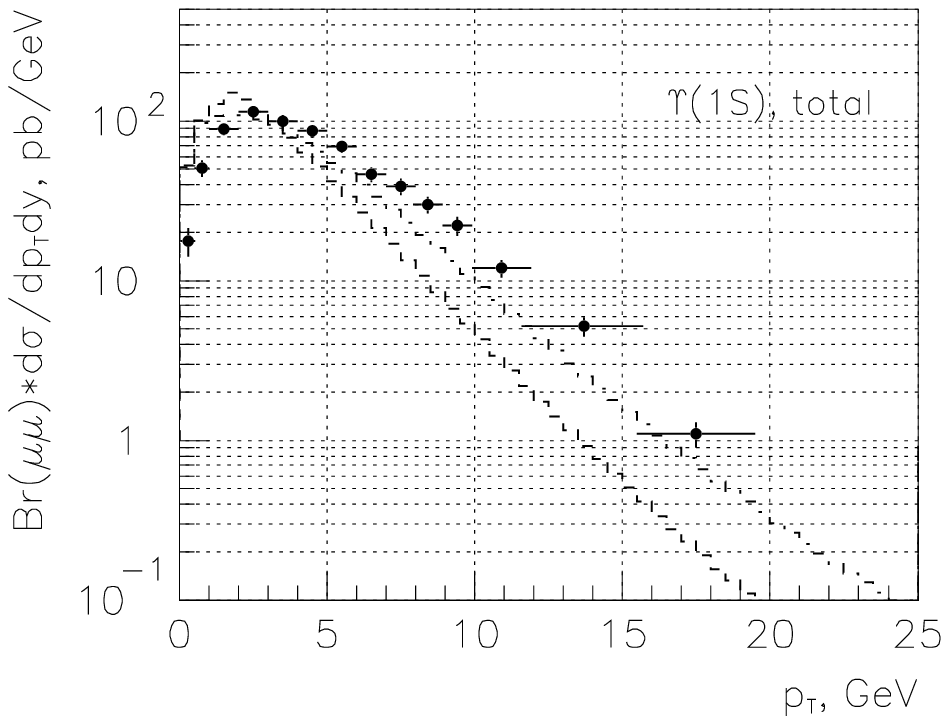}
\end{minipage}\hspace*{2mm}
\begin{minipage}[h]{.49\textwidth}
\includegraphics[height=5.0cm,width=6.0cm,angle=0]{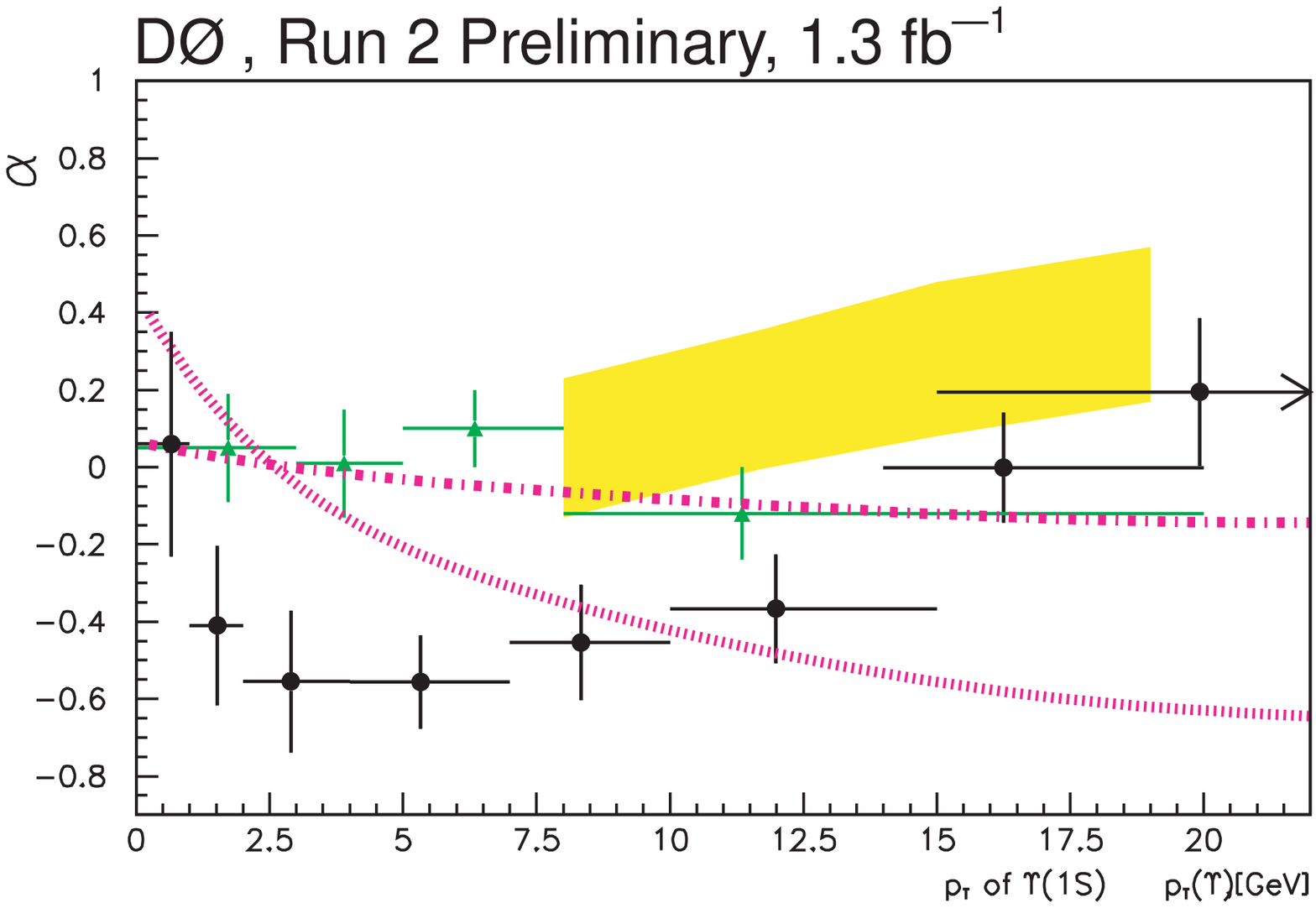}
\end{minipage}
\vspace*{-5mm}
\vspace*{-5mm}
\begin{center}
\caption[*] {Predictions on the production of {$\Upsilon (1S)$} mesons at the
Tevatron: the double differential cross section (left panel), the spin alignment 
parameter (right panel) as function of $p_T$. }
\end{center}
\end{figure}

The production of \U mesons in $pp$ collisions can proceed via either 
direct gluon-gluon fusion or the production of $P$-wave states $\chi_b$
followed by their radiative decays $\chi_b{\to}\u{+}\g$.
The direct mechanism corresponds to the partonic subprocess
$g+g\to\u+g$
which includes the emission of an additional hard gluon in the final 
state.
The production of $P$-wave mesons is given by
$g+g\to\chi_b$
only. All  essential parameters were taken from our previuos paper~\cite{bz}.
Fig. 1 shows the comparison of our results for 
the  \U (1S) meson production
with the D0 experimental data~\cite{D0cs, Kuz}.

 The calculations presented here
are also valid for the $\Upsilon(3S)$ state, except the lower
total cross section
(by an approximate factor of 1/3) because of the correspondingly 
lower
value of the wave function $|\Psi_{\u(3S)}(0)|^2=0.13$ GeV$^3$.
The state \U (3S) is producted by the purely direct production 
mechanism and the prediction on the spin alignment parameter 
$\alpha$ becomes less uncertain.

\begin{figure}[!thb]
\vspace*{10.0cm}
\begin{center}
\includegraphics{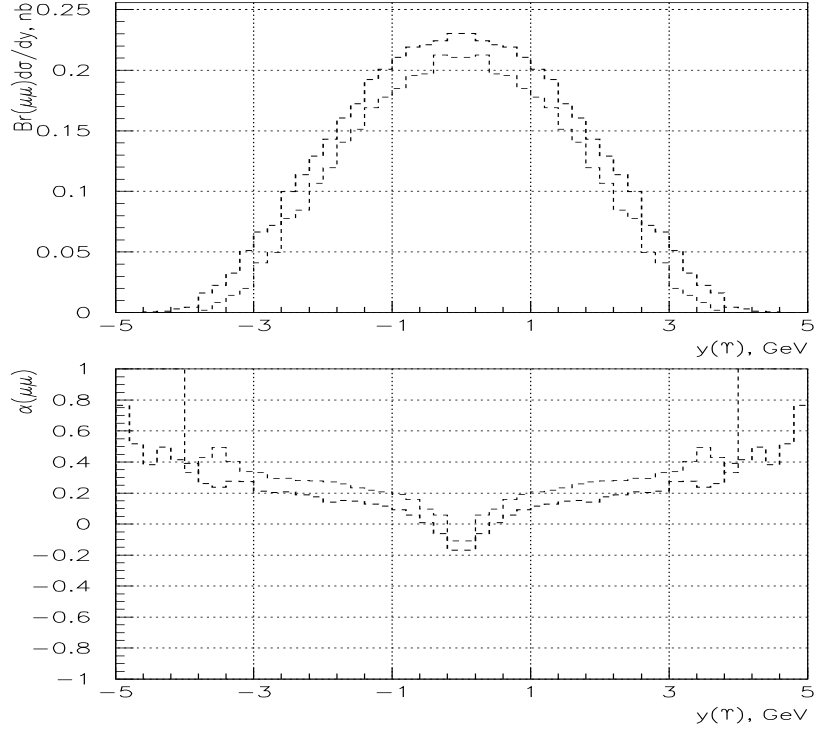}
\caption[*]{Predictions on the production of $\Upsilon (1S)$ mesons at the
Tevatron: the differential cross section and $\alpha$ parameter as function
of $y$.}
\end{center}
\label{fig1}
\end{figure}
The polarization state of a vector meson is characterized by the spin
alignment parameter $\alpha$ which is defined as a function of any
kinematic variable as
\begin{equation}\label{alpha}
 \alpha(\p)=(d\sigma/d\p -3d\sigma_L/d\p)/(d\sigma/d\p +d\sigma_L/d\p),
\end{equation}
where $\sigma$ is the reaction cross section and $\sigma_L$ is the part 
of cross section corresponding to mesons with longitudinal polarization
(zero helicity state). The limiting values $\alpha=1$ and $\alpha=-1$
refer to the totally transverse and totally longitudinal polarizations.
We will be interested in the behavior of $\alpha$ as a function of the
\U transverse momentum: $\p\equiv |{\mathbf p}_{T}|$. 
The experimental definition of $\alpha$ is based on measuring the
angular distributions of the decay leptons
\begin{equation}\label{dgamma}
d\Gamma(\u{\to}\mu^+\mu^-)/d\cos\theta\sim 1+\alpha\cos^2\theta,
\end{equation}
where $\theta$ is the polar angle of the final state muon measured in 
the decaying meson rest frame. Fig. 1 (right panel) shows the comparison of our results for the  \U (1S) meson polarization
with the D0 experimental data~\cite{Kuz}.
 The integration limits over rapidity were adjusted to the experimental
acceptances of D0 ($|y_\u|<0.6$) at the Tevatron. The yellow band in
Fig. 1 corresponds the NRQCD predictions, where the 
strong transverse polarization is connected with the gluon 
fragmentation mechanism.
 
In Fig. 2 we show the rapidity distribution for  \U (1S)
production in more wide rigion than the D0 experimental data, and
behaviour of the alignment parameter $\alpha$ as fuction of
rapidity. We see that  $\alpha$ becomes positive at large $y$.
Therefore we propose to measure the double differential cross
sections of quarkonium productions. 


In summary we have considered the production of \U mesons in high energy $pp$
collisions in the \ktf approach and compred the predictions on the 
spin alignment parameter $\alpha(p_T)$ with  new the D0 experimental data. We point out 
that the purest probe is provided by the polarization of $\Upsilon(3S)$ 
mesons.
We proposed to measure the double cross section of quarkonium 
production and the spin aligment parameter in more wide 
kinematical region on the rapidity also.

\end{document}